\documentclass{sig-alternate-2013}
\newcommand{\squeezeup}{\vspace{-2.5mm}} %% thanks http://tex.stackexchange.com/questions/60477/remove-space-after-figure-and-before-text
\permission{Copyright is held by the International World Wide Web Conference Committee (IW3C2). IW3C2 reserves the right to provide a hyperlink to the author's site if the Material is used in electronic media.}
\conferenceinfo{WWW'14 Companion,}{April 7--11, 2014, Seoul, Korea.} 
\copyrightetc{ACM \the\acmcopyr}
\crdata{978-1-4503-2745-9/14/04. \\
http://dx.doi.org/10.1145/2567948.2579272}

\usepackage{graphicx}
\usepackage{caption}
\usepackage{subcaption}

\begin{document}

\title{On the Ground Validation of Online Diagnosis with Twitter and Medical Records}


\numberofauthors{5} %  in this sample file, there are a *total*
\author{
\alignauthor
Todd Bodnar\titlenote{Corresponding author}\\
       \affaddr{Pennsylvania State University}\\
\affaddr{Center for Infectious Disease \\Dynamics and Department \\of Biology}\\
 %      \affaddr{University Park, PA 16802}\\
       \email{ToddBodnar@gmail.com}
\alignauthor
Victoria C Barclay\\
\affaddr{Pennsylvania State University}\\
\affaddr{Center for Infectious Disease \\Dynamics and Department \\of Biology}\\
 %      \affaddr{University Park, PA 16802}\\
       \email{vickistar50@gmail.com}
\and 
\alignauthor
Nilam Ram\\
\affaddr{Pennsylvania State University}
\affaddr{Department of Human \\Development and \\ Family Studies}
\email{nilam.ram@psu.edu}
\alignauthor
Conrad S Tucker\\
\affaddr{Pennsylvania State University}
\affaddr{Department of Engineering \\Design and Industrial and \\Manufacturing Engineering}
\email{conrad.tucker@psu.edu}
\alignauthor
Marcel Salath\'e\\
       \affaddr{Pennsylvania State University}\\
\affaddr{Center for Infectious Disease \\Dynamics and Department \\of Biology}\\
%       \affaddr{University Park, PA 16802}\\
       \email{salathe@psu.edu}
}

\maketitle
\begin{abstract}
Social media has been considered as a data source for tracking disease.  However, most analyses are based on models that prioritize strong correlation with population-level disease rates over determining whether or not specific individual users are actually sick. Taking a different approach, we develop a novel system for social-media based disease detection at the individual level using a sample of professionally diagnosed individuals. Specifically, we develop a system for making an accurate influenza diagnosis based on an individual's publicly available Twitter data. We find that about half (\(17/35 = 48.57\%\)) of the users in our sample that were sick explicitly discuss their disease on Twitter. By developing a meta classifier that combines text analysis, anomaly detection, and social network analysis, we are able to diagnose an individual with greater than 99\% accuracy even if she does not discuss her health.
\end{abstract}

%\category{I.6.4}{Simulation and Modeling}{Model Validation and Analysis}
%A category including the fourth, optional field follows...
%\category{D.2.8}{Software Engineering}{Metrics}[complexity measures, performance measures]


 % NOT required for Proceedings

\section{Introduction}
%Digital epidemiology, datamining Internet records to approach epidemological questions in novel ways, has recently been proposed as an alternative to traditional disease surveillance methods[]. 

Disease surveillance systems -- which traditionally rely on reports from medical practitioners -- are an important part of disease control. However, these traditional surveillance systems are often costly and slow to respond \cite{Chan2010,Heymann:2001,Salathe:2012ez}.  The widespread adoption of the Internet by the general public has provided opportunities for the development of novel disease surveillance methods. Compared to traditional systems, where data is provided by medical diagnosis, these new systems provide either semi-automatic -- through long term self reporting systems \cite{Marquet:2005tb,VanNoort:2007uk} -- or fully automatic -- through data mining search queries or social media \cite{Bodnar:2013we,Butler:2013uh,Culotta:2010hx,Goel:2010jf,Olson:2013bo} -- disease surveillance. While these methods are cheaper, faster and cover a larger number of individuals than traditional systems, one can be less confident about their results than the results from a system based on professional diagnosis. In this paper, we develop a system that performs long term surveillance on Twitter users with classifiers trained on professionally diagnosed data that combines the advantages of all three of these systems.

Previous work with data mining social media has focused on methods to replicate the patterns found in traditional surveillance networks \cite{Bodnar:2013we,Culotta:2010hx,Goel:2010jf}. However, these methods have several limitations. First, they generally do not differentiate between an individual with an illness and an individual that is worried about an illness; which may have resulted in a predicted influenza rate that was much higher then the actual 2013 influenza rate \cite{Bodnar:2013we,Butler:2013uh,Lamb:2013to,Olson:2013bo}. Second, these methods cannot be extended to areas without a previous surveillance network to train the model. Finally, these methods are fundamentally incapable of detecting diseases that do not show strong spatial-temporal patterns such as mental illness, obesity or Parkinson's disease. Instead of top-down methods to measure levels of disease in a population, we approach this problem from the bottom-up. This addresses all three of these issues: we only diagnose individuals that are likely to have the disease, and not just interested in the disease; we do not require previous data when applying these methods to new problems or locations; and these methods can easily generalize to diseases that do not show strong spatial or temporal patterns because we focus on an individual level.

Participatory systems, such as InfluenzaNet or Flu Near You, use self-reported symptoms to diagnose an individual and also work from a bottom-up approach \cite{Marquet:2005tb,VanNoort:2007uk}. These systems have the potential to be better than traditional surveillance systems because they update in near-real-time and can detect cases even when the user has not gone to their doctor. These systems require the user to sign up which allows for long term studies which are not normally able to be done with Tweets or search queries. However this reduces the number of users studied compared to data mining approaches. For example, Flu Near You had a total of 9,456 users report during the week ending on 29 December 2013. Marquet et al. \cite{Marquet:2005tb} have shown a large drop out rate with only 53\% of users participating for five or more weeks. While this amount of data is sufficient for many purposes, a system based on Twitter's millions of active users would open the door to more applications.

%Digital epidemiology \(\to\) novel disease detection mechanisms.
%
%Validation of this idea is important, but not done.
%
%Pull med info of individuals professionally diagnosed with ILI and their twitter accts. Compare old methods. Suggest some new things.

We develop such a system as follows. In section 2 we describe the collection of an individual's professional diagnoses of influenza and the collection of their Twitter information. In section 3 we consider extracting textual information from Tweets as a method for diagnosing influenza. Previous work has focused on this area. Additionally, we consider other methods for detection. In section 4 we consider anomalies in a user's Tweeting behavior as a signal for diagnosing influenza. In section 5 we extend these methods to other users on a person's social network to diagnose the original person. In section 6 we aggregate the results of the previous classifiers to develop a more accurate meta-classifier.

%\section{Related Work}

%People with issues \cite{Bodnar:2013we,Butler:2013uh,Lamb:2013to} also plos paper!

%Most work to this point considers finding messages in tweets (i.e. ``I'm sick'') or in keyword frequencies. 

%Keyword \cite{Culotta:2010hx,Goel:2010jf}
\squeezeup
%Tweet classification \cite{Culotta:2010hx,Lamb:2013to,Salathe:2011gr}
\begin{figure} [h]
\centering
\includegraphics[width=.45\textwidth]{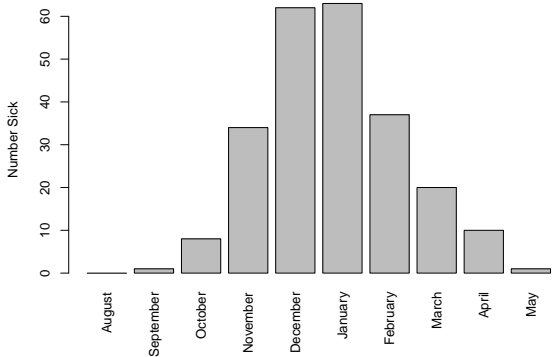}
\caption{The professionally diagnosed Influenza cases during the 2012-2013 season in our sample.}
\label{fig:flu_rate}
\end{figure}
\squeezeup
\section{Data Collection}
\subsection{Medical Records}
We received information from the Pennsylvania State University's Health Services about 104 individuals that were diagnosed with influenza by a medical professional during the 2012-2013 Influenza season. Due to privacy concerns, we were limited to knowing which month an individual was diagnosed (see figure \ref{fig:flu_rate}).  For comparison, we also obtained information from 122 individuals that were \emph{not} diagnosed with influenza during this time. The participants were mostly students (\(72\%\) were between the ages of 18 and 22) and slightly more female than expected (\(133 / 226 \approx 58.8\%\).) Data collection was approved through the Pennsylvania State University's IRB (approval \#41345.) Twitter handles were available for 119 of these individuals.

\subsection{Twitter Records}
While we received a total of 119 Twitter accounts, 15 were discarded because the associated accounts were either non-existent, banned or private. For each of the remaining 104 accounts, we pulled their profile information, their friends and followers information, their most recent 3000 tweets, and their friends' and followers' profiles and tweets. Some users did not tweet during the month that they were sick; we kept those accounts as part of the control group. We were limited to the most recent 3000 tweets by Twitter's time line query, but this only effected two accounts -- both of which posted multiple times per hour and were thrown out because we could only look back a few days.

We collected data through the Twitter API. Tweets, profile and follower information queries have separate rate limits and were collected in parallel. Since users continued to Tweet during data collection, each account was queried no more than once every three days for new Tweets. When all accounts could not be queried due to rate limiting, the accounts that had been queried the least recently were updated. Additionally, the 104 seed accounts collected above were given higher priority over their friends and followers. In total, we collected 37,599 tweets from the seed accounts and 30,950,958 tweets from 913,082 accounts that they either followed or were followed by.

%select count(distinct(d_net.user)) from (select distinct(user) as user from tweets_network) as d_net left join tweets on d_net.user = tweets.user where tweets.user is null;

%\section{Signal Detection}
\section{Text Based Signals}
\label{sec:text_analysis}

In this section, we consider diagnosis based on the content of a user's tweets. Such analysis can be approached by keyword analysis, where the presence of absence of a keyword predicts disease, or through text classification, where the tweets are classified as being about disease or not about disease. We begin by dividing the tweets into two sets: tweets that were posted the same month that a user was sick and tweets that were posted other times. We find a total of 1609 tweets from 35 users in the first category.

\begin{table}[h]
\centering
\begin{tabular}{|c|c|c|c|} \hline
Word& Total &Odds Ratio & Significance\ \\ \hline
flu&25&40.14& \textless 0.0001 \ \\ \hline 
influenza&1&0.00&0.8325\ \\ \hline 
sick&128&5.22& \textless 0.0001 \ \\ \hline 
cough&18&4.48&0.0094\ \\ \hline 
cold&82&1.45&0.4154\ \\ \hline 
medicin&9&11.20& \textless 0.0001 \ \\ \hline 
fever&13&26.20& \textless 0.0001 \ \\ \hline 

\end{tabular}
\caption{Probability of keywords being Tweeted by a user during the month that he or she was diagnosed with influenza.}
\label{tab:tweet_keyword_expert_results}
\end{table}

\begin{figure} [h]
\centering
\begin{subfigure}[b]{.2\textwidth}
\includegraphics[width=\textwidth]{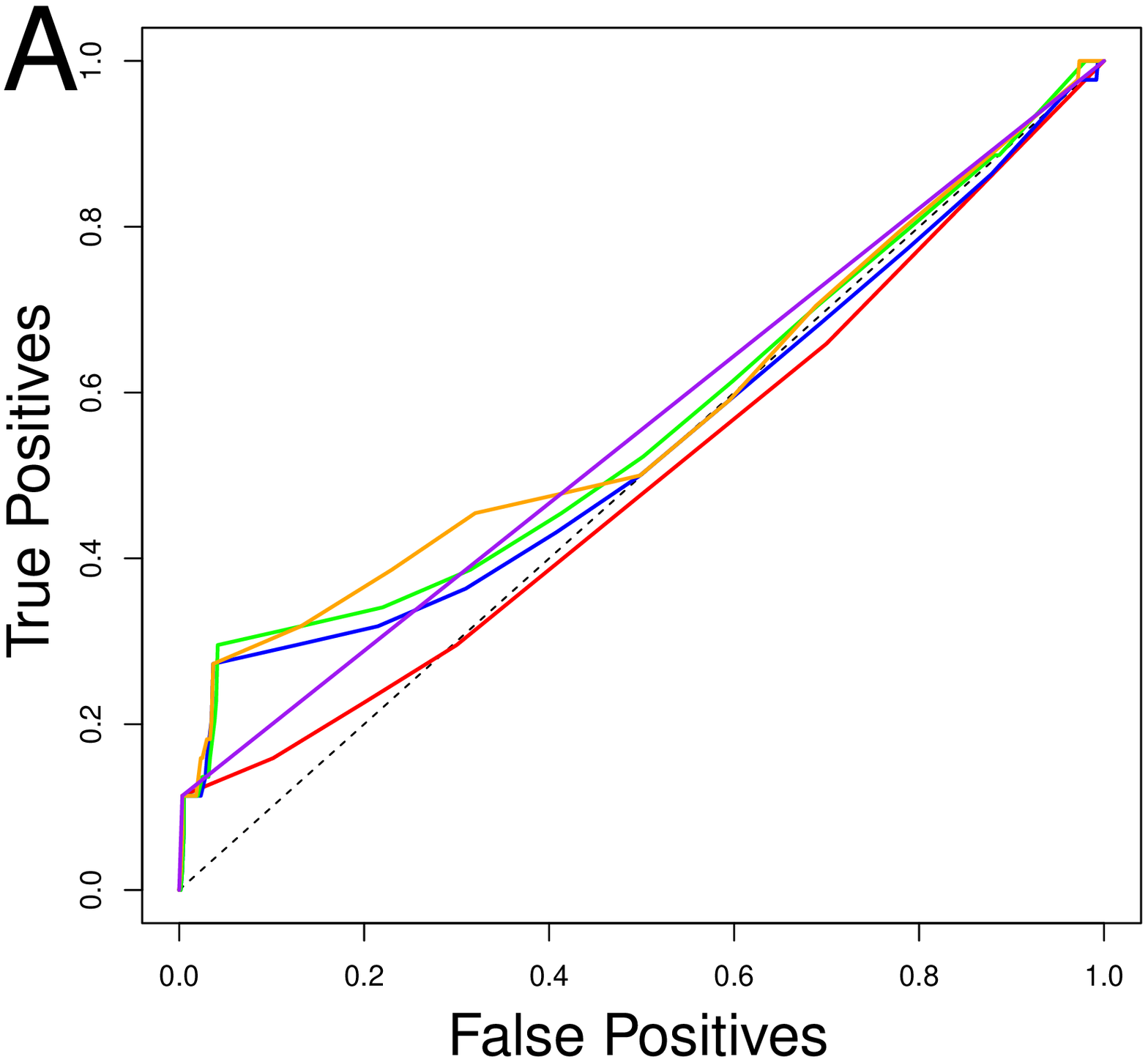}
\end{subfigure}
\begin{subfigure}[b]{.2\textwidth}
\includegraphics[width=\textwidth]{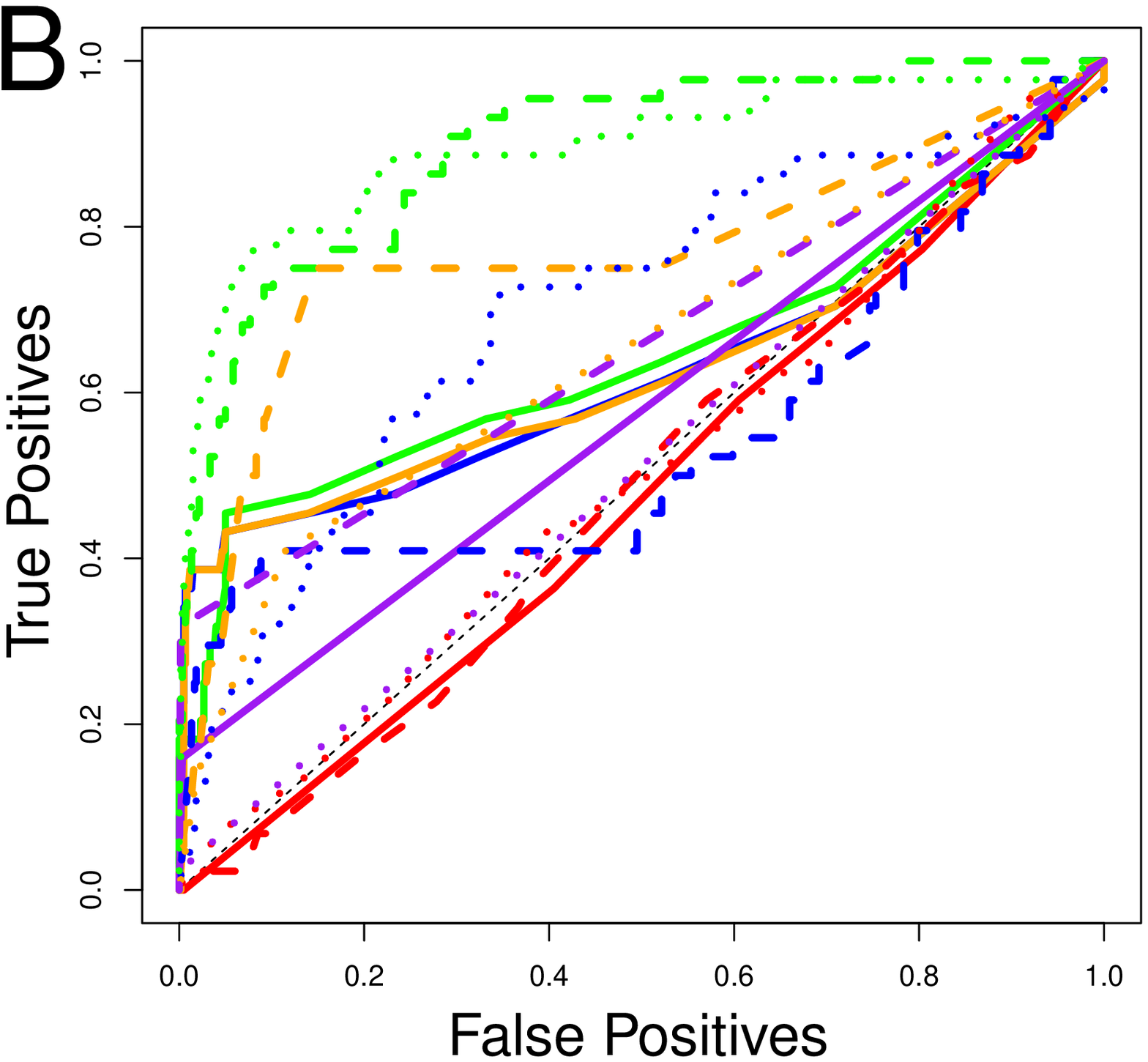}
\end{subfigure}
\begin{subfigure}[b]{.45\textwidth}
\includegraphics[width=\textwidth]{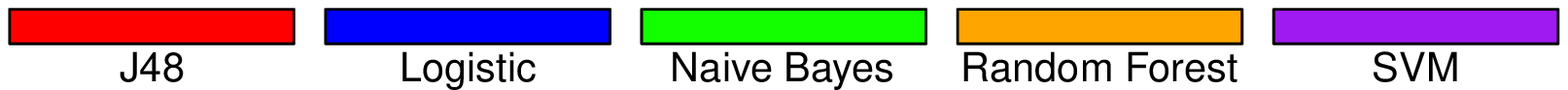}
\end{subfigure}
\caption{The ROC of classifiers that use hand chosen keywords (a) and algorithmically chosen keywords (b) to determine if an individual is ill. The top 10 (solid line), 100 (dashed line) and 1000 (dotted line) were selected as the features.}
\label{fig:roc_keyword}
\end{figure}

First, we use the occurrence or absence of keywords as features for classification. A set of keywords are defined that are possibly signals of influenza. We chose \{flu, influenza, sick, cough, cold, medicine, fever\} as our set of keywords. These keywords include the names and symptoms of the illness in addition to ``medicine'' and serve as a set of keywords that may have been chosen by a domain expert. We use Fisher's exact test to compare keyword occurrence in months when the user is sick or not sick and find a significant effect for six of the seven keywords (See table \ref{tab:tweet_keyword_expert_results}). Additionally, we try algorithmically selecting keywords by first finding the 12,393 most common keywords in the data set.  We then rank them based off of information gain on predicting influenza and choose the top 10, 100 or 1000 keywords from the list. In all of these cases, we pre-process the data by tokenizing the text on spaces, tabs and line breaks and the characters ``.,;':"()?!/\textbackslash '', remove stop words\footnote{Stop words were taken from Weka's stop list version 3.7.10.}, perform Porter stemming \cite{Porter:1980dd}  and convert the text to lower case. We use Naive Bayes, random forest, J48 (a Java implementation of C4.5), logistic regression and support vector machines to classify a user as being sick in a given month or not (see figure \ref{fig:roc_keyword}).

Second, we consider analysing the content of a tweet's text for messages giving hints about being sick such as ``another doctor's appointment Wednesday ... have to \#treatmyflu'' or ``I didn't realize how bad it feels to have the flu, should have gotten a flu shot\footnote{These examples are based off of real tweets, but changed to keep our participants anonymous.}'' that would not be detected through simple bag-of-words techniques. Computational approaches for natural language processing are available. However, because our dataset is relatively small, we use a `human' classifier by hand rating all 1609 tweets that were posted by individuals during the time of their illness. We also sample a randomly selected set of 1609 tweets from times when the users did not have influenza as a control. We find 58 tweets from 17 (\(17/35 = 48.57\%\))  individuals in our study that are about the user being sick. We also find zero tweets about the user having influenza during times when they did \emph{not} have influenza. Because humans are very good at extracting information from text, hand rating tweets allows for an approximately 100.0\% accurate classification, although it clearly does not scale well. Extracting information from text using machine learning is a complex problem where finding solutions that perform as well as humans is rare. Thus, the human classifier gives us an upper limit to the accuracy of a health monitoring system based off of tweet classification (see table \ref{tab:tweet_classified_confusion}.)

\begin{table}
\centering
\begin{tabular}{|c|c|c|} \hline
Sick&Not Sick&\ \\ \hline
17 & 18 & Sick\\ \hline
0 & 66  & Not Sick\\
\hline\end{tabular}
\caption{Confusion matrix of a Tweet-Classification based diagnosis system. Rows are of true values, columns are of predicted values.}
\label{tab:tweet_classified_confusion}
\end{table}

\section{Frequency Based Signals}

In addition to illness affecting the content of individuals' tweets, it is likely that illness also affects the rate at which individuals tweet. To detect this, we perform one-dimensional anomaly detection on each user's monthly tweeting rate as follows. First, we calculate the number of tweets in each month in the study period and discard any months where the user tweets less than ten times. This avoids issues caused by the user starting or stopping their use of Twitter. We then calculate the z-score of the tweeting rate of the month that the user is ill by
\begin{equation}
z = \frac{|x - \bar{x}|}{\hat{s}}
\end{equation}

Where \(\bar{x}\) and \(\hat{s}\) are the estimated mean and standard deviation of the user's tweeting rate for each month during the study \cite{Grubs:1969ab}. We repeat this process for months when the user is not sick. We then classify the user as sick if \(z > 1.411\) where \(1.411\) was chosen through leave one out cross validation. We find a significant difference between the z-scores for months when a user had influenza and months when the user did not (\(p = 0.01303\), two-sample Kolmogorov-Smirnov test). Most of the time individuals are not sick (219 / 258 = 84.88\% of the months), resulting in a highly biased sample. Thus we optimize based on the \(F_1\) score instead of accuracy. The optimal z-score cutoff results in  an area under the ROC curve of .6218 and \(F_1= 35.0\%\). (See table \ref{tab:tweet_anomaly_confusion}.) 

%\begin{figure} %need to redo
%\centering
%\includegraphics[width=0.5\textwidth]{figs/meanFrequencies.eps}
%\caption{The frequency of tweeting behaviour of individuals in the months before, during and after an illness. Users significantly (check) decrease their rate of tweeting during the time that they had influenza. Dashed lines indicate the mean rate for the three months before / after the illness. (Todo: check significance)}
%\label{fig:mean_freq}
%\end{figure}

\begin{table}[h]
\centering
\begin{tabular}{|c|c|c|} \hline
Sick&Not Sick&\ \\ \hline
14 & 25 & Sick\\ \hline
27 & 192 & Not Sick\\
\hline\end{tabular}
\caption{Confusion matrix of the classifier based on anomalous tweeting rates. Rows are of true values, columns are of predicted values.}
\label{tab:tweet_anomaly_confusion}
\end{table}

\section{Network Based Signals}

\begin{figure} [h]
\centering
\begin{subfigure}[b]{.2\textwidth}
\includegraphics[width=\textwidth]{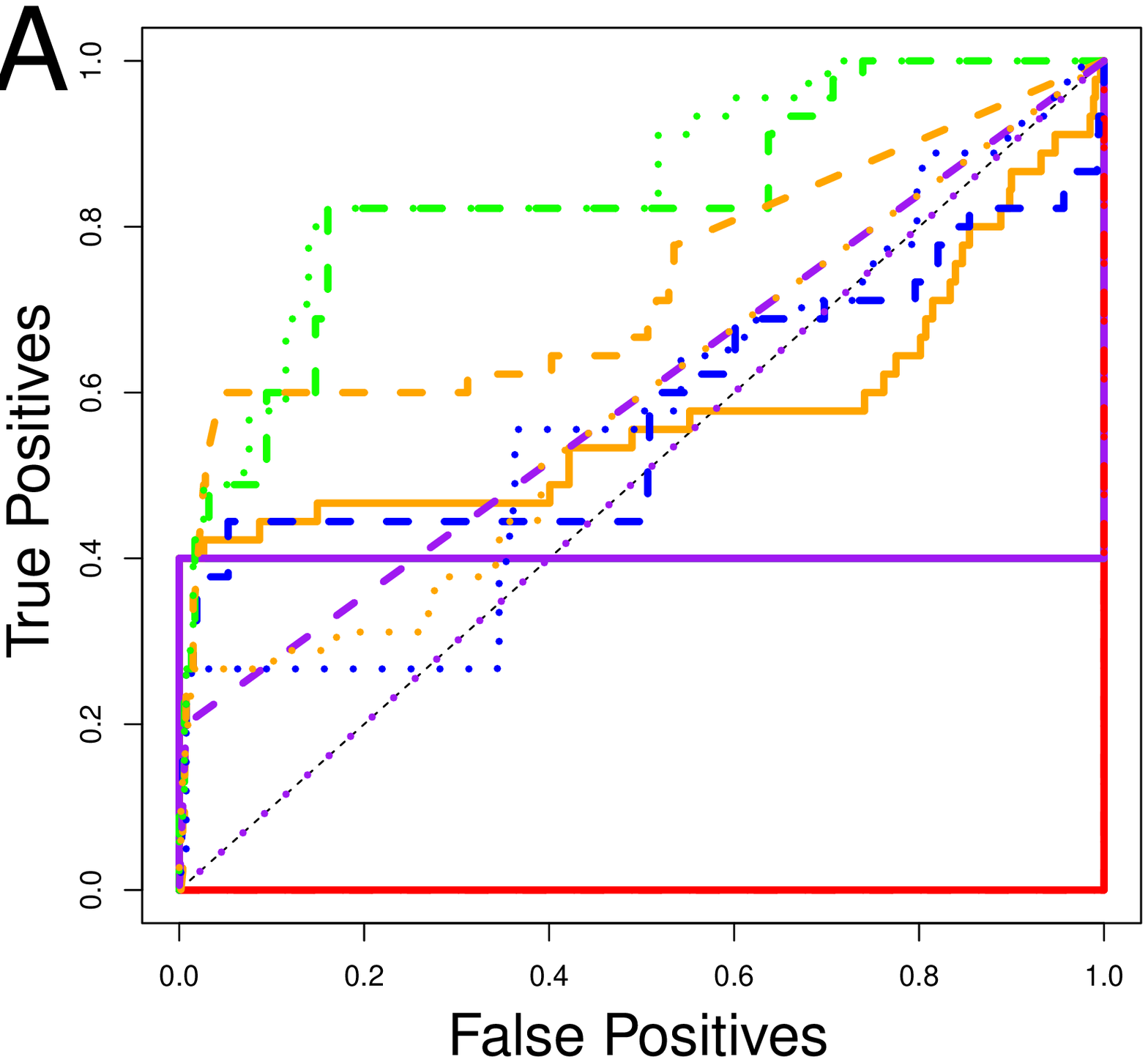}
\end{subfigure}
\begin{subfigure}[b]{.2\textwidth}
\includegraphics[width=\textwidth]{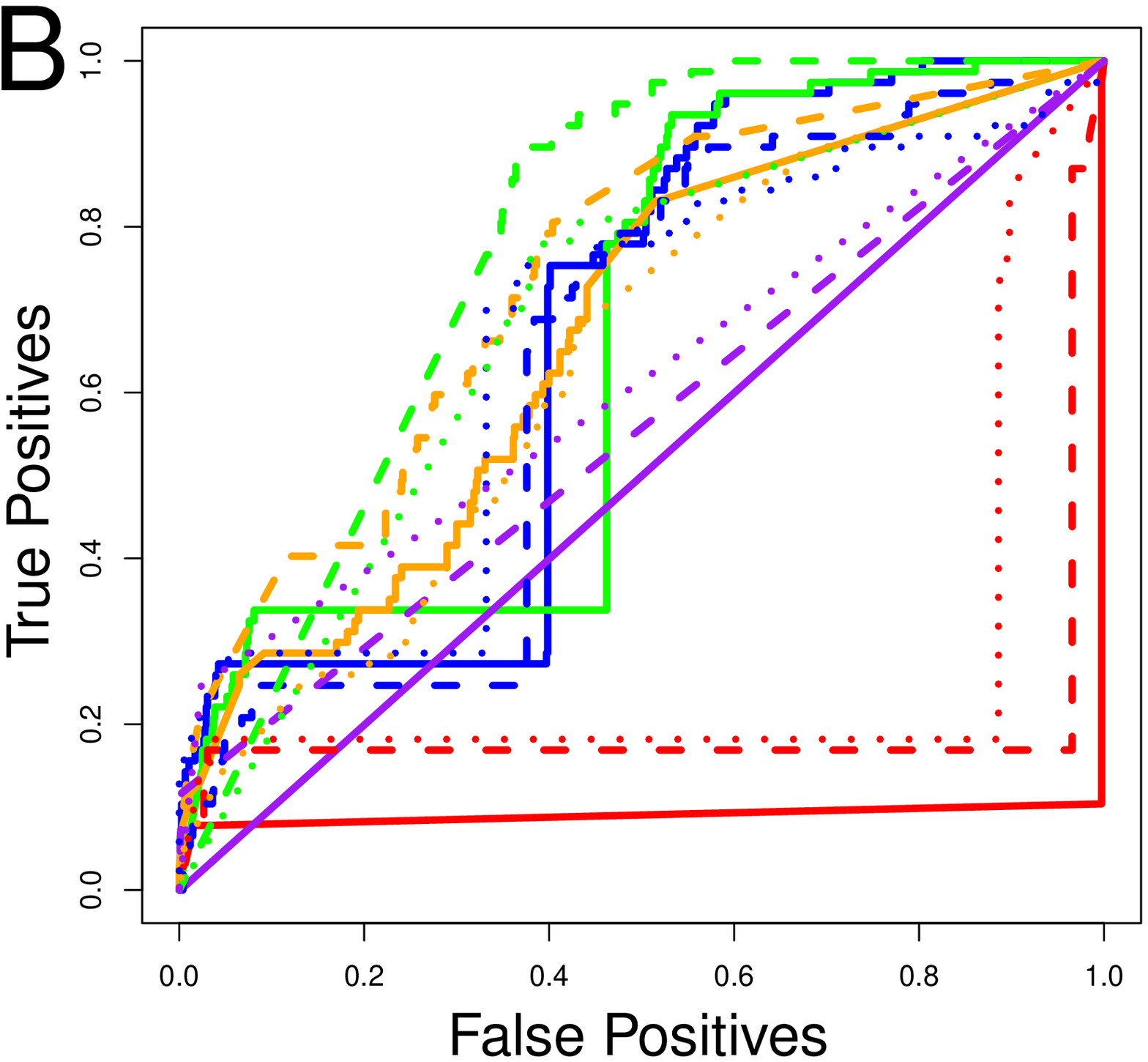}
\end{subfigure}
\begin{subfigure}[b]{.45\textwidth}
\includegraphics[width=\textwidth]{figs/keyword_legend.eps}
\end{subfigure}
\caption{The ROC of classifiers based off Tweets from (a) accounts that follow a user and (b) accounts that a user follows. Line coloring and style are equivalent to figure \ref{fig:roc_keyword}.}
\label{fig:roc_network}
\end{figure}

Even if a user is not currently active on Twitter, users on her social network may give clues to her health status. Twitter's social network is one directional, allowing for users to follow other users without the other users having to follow them back. Accounts that follow a user are referred to as her `followers,' and accounts that a user follow are referred to as her `friends.' We consider all text that a user's friends or followers tweeted and perform keyword analysis. The analysis was performed the same way as we analyzed the user's tweets in section \ref{sec:text_analysis}, except we normalize the counts here by the total number of characters her followers or friends tweeted. This controls for the number and activity of a users friends or followers, which should not have an effect on her health status. We find that most of the tested classifiers are able to detect a signal in both the user's followers' and friends' streams (see figure \ref{fig:roc_network}.)

%follower 12393
%friend 11595
%best = nb_1000 roc=.8641

We further analyse the strength of these classifiers by building each classifier using 10 fold cross validation and calculate their performance by measuring area under the ROC curve. We repeat this 100 times to generate a distribution of each classifier's performance. We then perform an analysis of variance test to examine the differences between the sources of data (followers or friends), the number of keywords used and the classifier's algorithm (see table \ref{tab:network_aov}.) We find that the choice in classifier and the length of the feature vector have a significant effect on performance. We find that classifiers that use tweets from accounts that follow the user are significantly better at diagnosing the user than classifiers that use tweets from accounts that the user follows. This may be because Twitter users often follow celebrities and news organizations -- and celebrities and news organizations rarely follow personal Twitter accounts --  which could introduce excess noise. 

\begin{table}[htp]
\begin{center}
\begin{tabular}{llllll}
\hline\hline
\multicolumn{1}{c}{ }&\multicolumn{1}{c}{Df}&\multicolumn{1}{c}{Sum Sq}&\multicolumn{1}{c}{F value}&\multicolumn{1}{c}{Pr(\textgreater F)}\tabularnewline
\hline
Source&1&107.16&1290.82&\textless \(2^{-16}\) \tabularnewline
Keyword Size&1&72.19&869.66&\textless \(2^{-16}\)\tabularnewline
Classifier&3&752.55&3021.61&\textless \(2^{-16}\)\tabularnewline
Residuals&109194&9602& & \tabularnewline
\hline
\end{tabular}
\end{center}
\caption{Results from an analysis of variance of the area under the ROC curve for classifiers based on tweets from an individual's social network. Factors are whether the data is from the user's friends or followers, the number of keywords chosen and the classifier. }
\label{tab:network_aov}
\end{table}
\squeezeup
\section{Meta Classifier}

So far we have considered five separate methods for detecting illness based off of a user's Twitter activity: hand-chosen keyword analysis, datamined keyword analysis, hand classified tweets, anomaly detection and network analysis. However, there is no reason that we cannot combine these methods to get a stronger signal. For example, while mining the user's text is the best of the five methods, she may stop tweeting while sick, which would be detected by the frequency-based anomaly classifier. Aggregating multiple classifiers by a `meta-classifier' has been shown to be an effective method for increasing classification accuracy \cite{Todorovski:2003hk,Frossyniotis:2004wx}.

\begin{table}
\centering
\begin{tabular}{|c|c|c|} \hline
Classifier&Area under ROC& Accuracy \\ \hline
AdaBoost & .9961 & 99.53\\ \hline
Bayesian & .9078 &  92.08\\ \hline
Decision Tree & .9877 & 99.22\\ \hline
Logit Boost & .9986 & 99.22\\ \hline
Weighted Voting & .9783 & 93.17\\ \hline
Baseline & .8544 & 89.72\\ 
\hline\end{tabular}
\caption{Performance of the meta classifiers. The presented baseline is the classifier based on datamined keywords -- the highest preforming individual classifier.}
\label{tab:meta_results}
\end{table}

\begin{figure} [h]
\centering
\begin{subfigure}[b]{.2\textwidth}
\includegraphics[width=\textwidth]{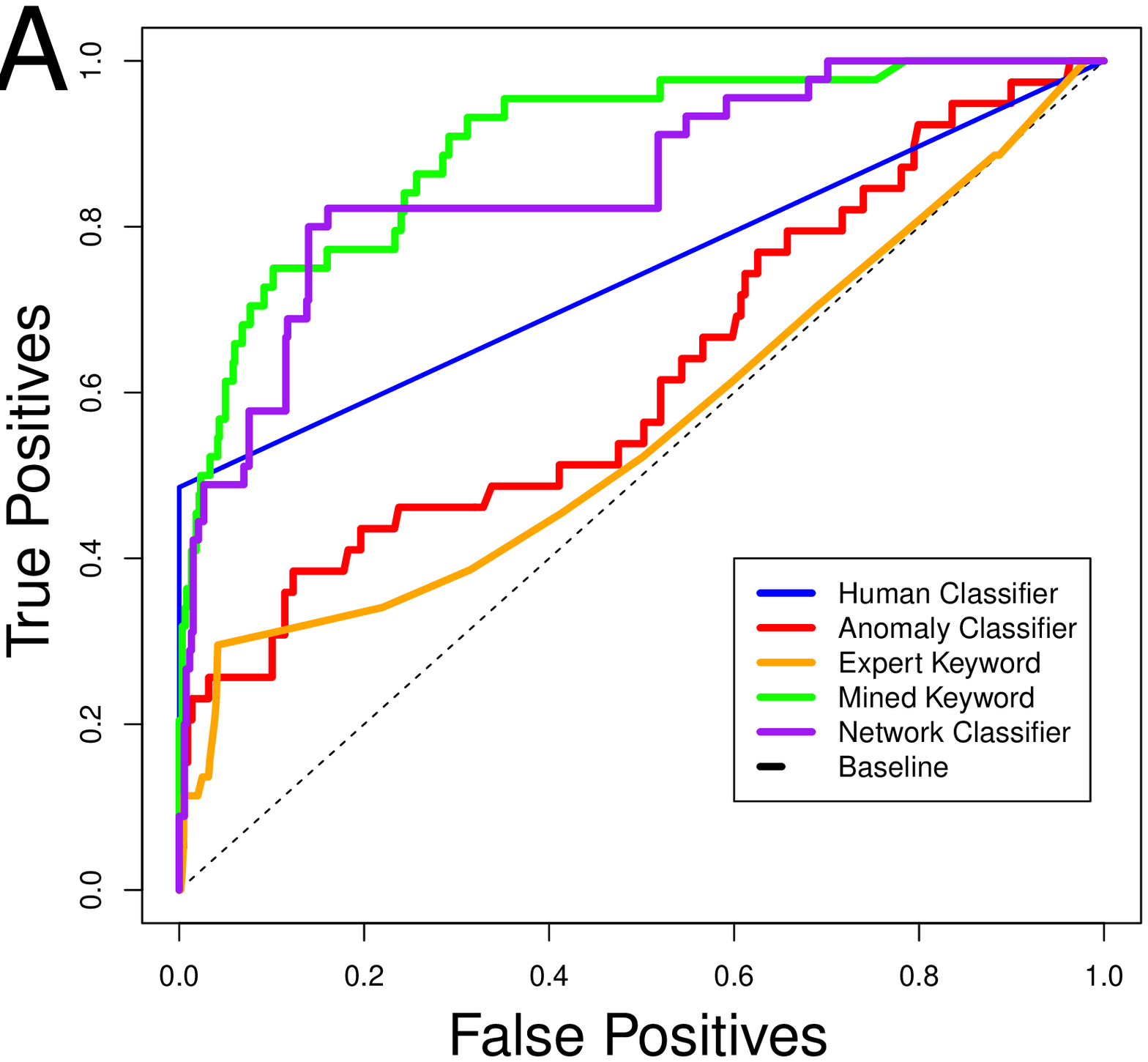}
\end{subfigure}
\begin{subfigure}[b]{.2\textwidth}
\includegraphics[width=\textwidth]{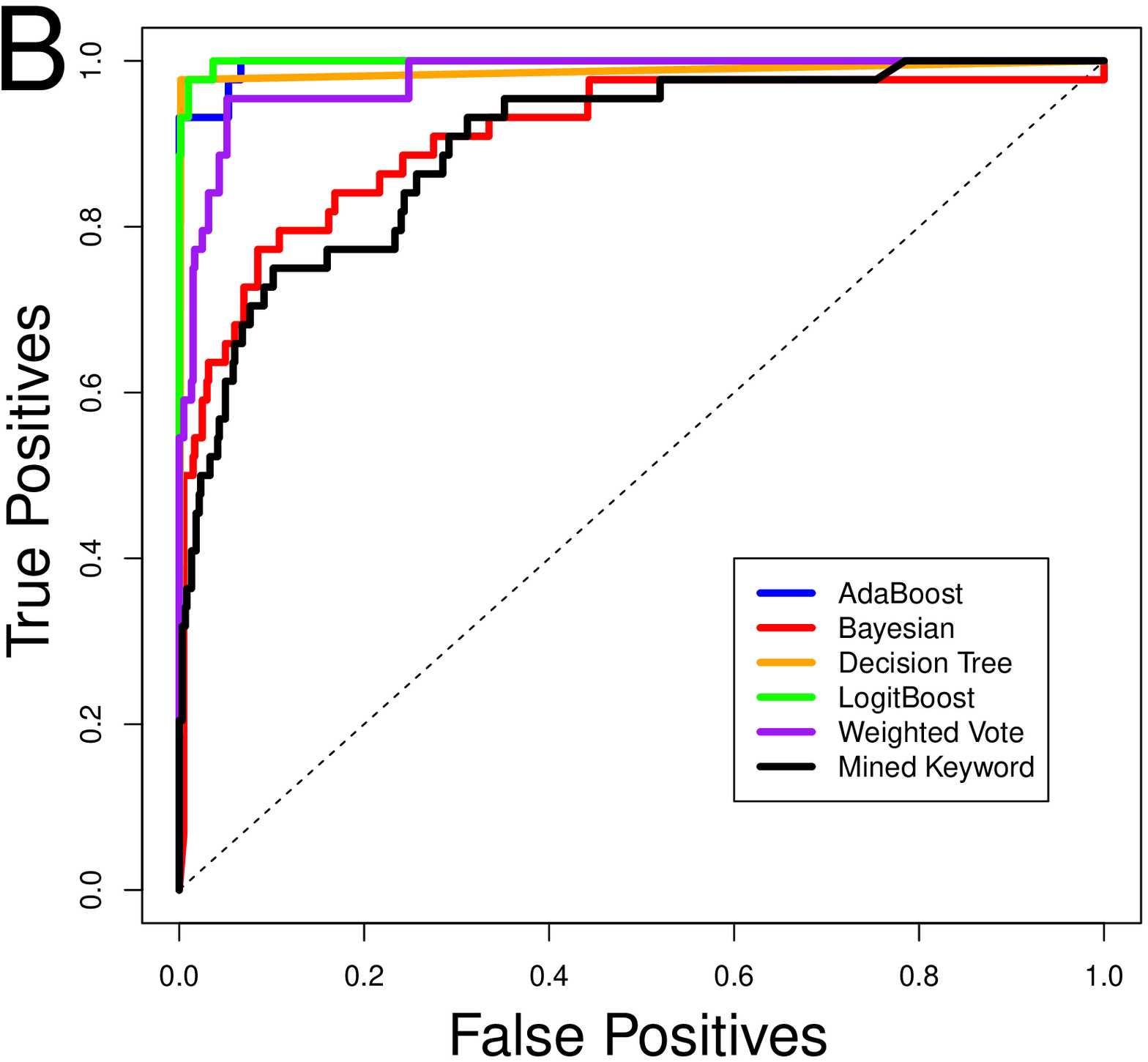}
\end{subfigure}
\caption{The accuracy of the previous classifiers (a) and the accuracy of various classifiers that use the previous classifier's results as features (b). }
\label{fig:roc_meta}
\end{figure}

We start by selecting the classifier from each of the previous five approaches that has the largest area under the ROC curve (see figure \ref{fig:roc_meta}.a.) We then use the predicted distributions from these classifiers as the feature vector for the meta classifier. We use Ada Boost, Bayesian classification, J48 decision trees, logit boost, and weighted voting to evaluate the meta-dataset. We then evaluate these methods with leave-one-out cross validation and see an increase in area under ROC and accuracy compared to the best individual classifier (see figure \ref{fig:roc_meta}.b.) We find that AdaBoost has the highest accuracy (99.53\%) and logit boost has the highest area under it's ROC curve with .9986 (see table \ref{tab:meta_results}.)

%adaboost, weighted voting, logistic, j48 decision trees, basian model, stacking, logit boost

\squeezeup

\section{Conclusions}

In this paper, we have shown that it is possible to diagnose an individual from her social media data with high accuracy. Computational approaches to aid in disease diagnosis has been approached before, however they have been developed with a medical setting in mind. That is, the question addressed was ``can we diagnose an individual based off data gathered from medical tests run on her?'' instead of ``can we diagnose an individual solely based off of publicly available social media data?''  While we focus on the relatively benign case of remotely reconstructing a confidential diagnosis of influenza, these methods could also be applied to stigmatized diseases, such as HIV, where being able to determine if an individual is HIV positive without her knowledge and with only her Twitter handle could result in serious social or economic effects. Half of the users explicitly stated that they were sick, and we were able to confidently determine illness in the other half of the cases through their data. It would seem that simply avoiding discussing an illness is not enough to hide one's health in the age of big data. 

%\section{Conclusions}
\begin{table}[!ht]
\centering
\begin{tabular}{|c|c|}\hline
Keyword&Ratio\ \\ \hline
flu &  34.424\\ \hline
health &  11.360\\ \hline
sick &  5.019\\ \hline
track & 10.952 \\ \hline
stud & 3.508 \\ \hline
asshol & 9.090 \\ \hline
ton & 9.090 \\ \hline
particip & 20.667 \\ \hline
salt & 20.667 \\ \hline
recov & 40.118 \\ \hline
fuck & 2.963 \\ \hline
sham & 13.64 \\ \hline
row & 10.180 \\ \hline
win & 2.947 \\ \hline
rt & 3.077 \\ \hline
  \end{tabular}
  \hspace{1em}
\begin{tabular}{|c|c|}\hline
cont. & \\ \hline
walk & 3.077 \\ \hline
childr & 6.820 \\ \hline
incred & 6.820 \\ \hline
meal & 6.820 \\ \hline
longer &  6.820 \\ \hline
succes &  26.765 \\ \hline
accis & 26.765 \\ \hline
holida & 26.765 \\ \hline
luv & 26.765 \\ \hline
oblig & 26.765 \\ \hline
path & 26.764 \\ \hline
pract & 26.764 \\ \hline
prayer & 26.765 \\ \hline
reserv & 26.765 \\ \hline
riot & 26.765 \\ 
\hline\end{tabular}
\caption{The thirty keyword stems with the highest positive predictive power ranked by significance. The Twitter API limits searches to at most thirty keywords. Ratio is calculated as the rate of occurrence when a user is sick over the rate when a user is not sick.}
\label{tab:thirty_best}
\end{table}

\section{Acknowledgments}

Marcel Salath\'e received funding through a Branco Weiss fellowship. We thank the Pennsylvania State University's Student Health Center for aid in collecting data. We thank Cosme Adrover Pacheco for his valuable comments on the paper. The machine learning tool set Weka version 3.7.10 was used in this paper \cite{Hall:2009ud}.

%\bibliographystyle{abbrv}
%\bibliography{library}  % sigproc.bib is the name of the Bibliography in this case
% ACM needs 'a single self-contained file'!
%\balancecolumns
%\balancecolumns
\squeezeup
\appendix
\section{Keyword Recommendations}
While our system should be trusted more than one based simply off of aggregated tweets, it is more computationally intensive than simply pulling data from a keyword stream. These systems require the user to select a specific set of keywords before data collection can begin. Keywords representing symptoms such as ``flu'', ``cough'', ``sore throat'', and ``headache'' are often chosen. We suggest  the thirty keywords  with the highest positive predictive value (see table \ref{tab:thirty_best}) be chosen as the parameters for a keyword stream. In addition to keywords related to symptoms (e.g. ``flu'' or ``sick'')  we also find keywords related to treatments (e.g. ``health,'' ``prayer'' or ``recovery'') and keywords related to negative mood (e.g. vulgarities) to be more commonly tweeted when a user is ill. 
 
\vspace{4pt}

\balancecolumns

% That's all folks!
\end{document}